\documentclass[vecphys]{svmult}

\usepackage{makeidx}         
\usepackage{graphicx}        
\usepackage{subfigure}
\usepackage{multicol}        
\usepackage[bottom]{footmisc}
\usepackage{amsmath}
\usepackage{amssymb}
\newcommand{\ham}{{\mathcal H}}

\makeindex             

\begin{document}

\title*{The Gonihedric Ising Model and Glassiness}
\author{D.A. Johnston\inst{1}, A. Lipowski\inst{2} and Ranasinghe P. K. C. Malmini\inst{3}}

\institute{Department of Mathematics and the Maxwell Institute for Mathematical Sciences,
    Heriot-Watt University,
	 Riccarton Edinburgh, EH14 4AS, Scotland,
\and
Faculty of Physics, Adam Mickiewicz University, 61-614 Pozna\'n, Poland
\and
                  Department of Mathematics,
University of Sri Jayewardenepura,
Gangodawila, Sri Lanka.}

%
%
\maketitle

\section{(Pre-)History of the Model}
The Gonihedric $3D$ Ising model is a lattice spin model in which planar Peierls 
boundaries between $+$ and $-$ spins can be created
at zero energy cost. Instead of weighting the {\it area}  of Peierls 
boundaries as the case for the usual $3D$ Ising model with nearest neighbour interactions, 
the {\it edges}, or ``bends'' in an interface are weighted, a 
concept which is related to the intrinsic curvature of the boundaries 
in the continuum. 

The model is a generalised Ising model living on a cubic $3D$ lattice with nearest neighbour, next to nearest-neighbour
and plaquette interactions. The ratio between the couplings of these three terms is fixed to a 
one parameter family which endows
the model with unusual properties both in  and out of equilibrium. Of particular interest
for the discussion here will be that the model manifests  all the indications of glassy behaviour
without any recourse to quenched disorder,
whilst still possessing a crystalline low temperature phase in equilibrium. 

In these notes we follow a roughly chronological order by first reviewing the background to the formulation of the model, before moving on to the elucidation of the equilibrium phase diagram by various means and 
then to the investigation of the non-equilibrium, glassy behaviour of the model. We apologize in advance
for our narrow focus on things Gonihedric at the expense of other lattice models with glassy behaviour,
since the aim is to concentrate on giving an overview of the Gonihedric Ising model in $3D$.

\begin{figure}
\centering
\includegraphics[height=7.5cm]{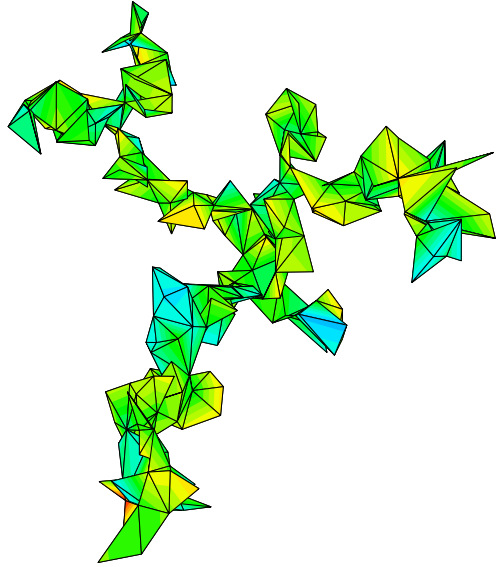}
%
%
\caption{A typical collapsed triangulated surface resulting from a simulation of the Gaussian Hamiltonian
in equ.\ref{pol}.}
\label{collapsed}       
\end{figure}

The model has an unusual genesis since it was originally introduced 
as a potential discretization of  string theory. The Nambu-Goto \cite{1} action (or Hamiltonian
in statistical mechanical language) in bosonic string theory is 
given by the area swept out by the string worldsheet as it moves through spacetime.
Directly discretizing euclideanized versions of this action produced ensembles of surfaces, {\it i.e.} string worldsheets,
which were dominated by collapsed and irregular configurations such as that in Fig.\ref{collapsed}
and which were unsuitable for 
taking any sort of continuum limit \cite{2}. This was presumably a reflection of the well-known tachyonic instability of the bosonic
string and therefore not entirely unexpected. The interest was in whether any straightforward modifications
of such Hamiltonians did allow a continuum limit, and hence some insight
into stable string theories in physical dimensions.

Discretizing the Polyakov \cite{3} form of the string action,
which is equivalent to the Nambu-Goto formulation,
by triangulating  an embedded  surface gives the simple Gaussian Hamiltonian
\begin{equation}
H = {1 \over 2}  \sum_{\langle ij \rangle} ( \vec X_i - \vec X_j )^2
\label{pol}
\end{equation}
where the vectors $\vec X_i$ live on the vertices of the triangulated surface and determine where the vertices
sit in the embedding space. The sum $ \langle i j  \rangle$ is carried out over the edges of the
triangulation and in principle one should also sum over different triangulations
of the surface
as a discretized version of the sum over metrics in the continuum string theory with the Polyakov action.
In any case,
the observed behaviour of surfaces turns out to be rather similar in ensembles  with and without this sum. 
One possibility for stabilising the surfaces produced by such Hamiltonians
is to add  an extrinsic curvature term as suggested first in the continuum by Polyakov and Kelinert \cite{4}, which acts to prevent the surfaces collapsing
\begin{equation}
H = {1 \over 2}  \sum_{\langle ij  \rangle} ( \vec X_i - \vec X_j )^2 +
\lambda \sum_{\Delta_i, \Delta_j} ( 1 - \vec n_i  \cdot \vec n_j )
\label{pol+n}
\end{equation}
where the $\vec n_i, \vec n_j$ are normals on adjacent triangles $\Delta_i, \Delta_j$.
The second term tends to align the normals on adjacent triangles
and hence to ``stiffen'' or flatten out the surface \cite{5}, at least 
over short length scales. 
\begin{figure}
\centering
\includegraphics[height=5.5cm]{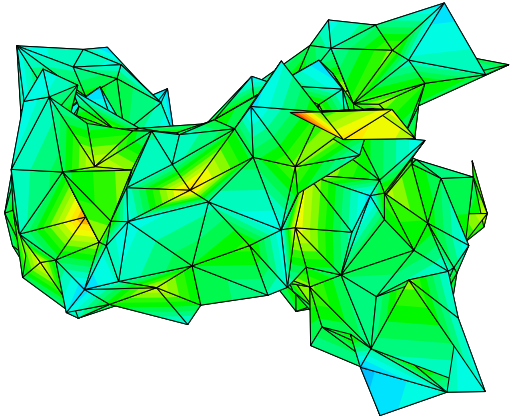}
%
%
\caption{A typical uncollapsed triangulated surface resulting from a simulation of the Gaussian 
plus extrinsic curvature Hamiltonian in equ.\ref{pol+n} with $\lambda = 1.1$.}
\label{uncollapsed}       
\end{figure}
The issue for Monte Carlo simulations is to determine whether this flattening is also effective
on macroscopic length scales.

\begin{figure}
\centering
\includegraphics[height=3.5cm]{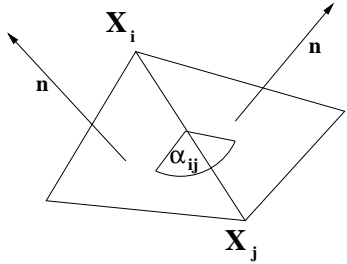}
%
%
\caption{Two adjacent triangles in a triangulation of a 
surface with a common edge $\langle ij \rangle$, showing the normals $\vec n$,
the co-ordinates of the endpoints $\vec X_{i,j}$ and the dihedral angle $\alpha_{ij}$}
\label{dihedral}       
\end{figure}

Another possibility, which also uses a geometrical concept and which has
some appealing features from the string theory point of view \cite{6}, is to take  an action proportional
to the  linear size of a surface -- a notion first introduced by Steiner \cite{6a}.
This can be discretized as
\begin{equation}
H = {1 \over 2} \sum_{\langle ij \rangle} | \vec X_i - \vec X_j | \; \theta (\alpha_{ij}),
\label{steiner}
\end{equation}
on triangulated surfaces, where
$\theta(\alpha_{ij}) = | \pi - \alpha_{ij} |$
where $\alpha_{ij}$ is the dihedral angle between the
embedded neighbouring triangles with a common link $\langle ij \rangle$. 
The $\theta (\alpha_{ij})$ term also acts to flatten out the surfaces \cite{6b}. The  elements
making up these various possible surface Hamiltonians are shown in  Fig.\ref{dihedral}.
The origin of the word Gonihedric is a combination of the Greek words gonia for angle (referring to the dihedral angle)
and hedra for base or face (referring to the adjacent triangles).

Savvidy and Wegner posed the question of how
to translate the essential features of such a Hamiltonian to a 
surface composed of   
plaquettes on a cubic (or hypercubic) lattice \cite{7}.
This was motivated by the observation that
it was possible to rewrite the Hamiltonian
for an ensemble of such surfaces 
by using the geometrical spin cluster boundaries 
of a suitable generalised Ising model with nearest neighbour, next to nearest-neighbour
and plaquette interactions to define the surfaces and their appropriate energies.
It is generally much easier to simulate an Ising spin model than a gas of surfaces
so one might expect some advantages, both numerical and analytical, from such a
reformulation of the system. 

On a fixed cubic lattice all the edge lengths $| \vec  X_i - \vec X_j |$
will be identical, so the statistical weight of a surface configuration will be entirely determined by
the $\theta(\alpha_{ij}) = | \pi - \alpha_{ij} |$ factors, where the $\alpha_{ij}$ are restricted to
multiples of $\pi / 2$ radians. In other words, the statistical weight of a plaquette surface configuration 
will be 
determined entirely by the number of bends and self-intersections it contains. There is no (bare)
weight for the area of a plaquette in the surfaces in contrast to the usual $3D$ Ising model with only nearest neighbour interactions.

The generalised Ising models which are used in the construction
are of the form
\begin{equation}
H = J_1  \sum_{\langle ij\rangle }\sigma_{i} \sigma_{j} + 
 J_2 \sum_{\langle \langle i,j\rangle \rangle }\sigma_{i} \sigma_{j} 
+ J_3 \sum_{[i,j,k,l]}\sigma_{i} \sigma_{j}\sigma_{k} \sigma_{l}.
\label{e0}
\end{equation}
where
the Hamiltonian contains nearest neighbour ($\langle i,j\rangle $),
next to nearest neighbour ($\langle \langle i,j\rangle \rangle $) and round a plaquette ($[i,j,k,l]$)
terms.
The couplings in such models can be related to the energy cost for a plaquettes  in the Peierls interface  $\beta_A$, a right-angled
bend between two such adjacent plaquettes, $\beta_C$, and the energy cost, $\beta_I$, for the intersection of four plaquettes having a link in common 
\begin{eqnarray}
\beta_A &=&  2 J_1 + 8 J_2 \nonumber \\
\beta_C &=&  2 J_3 - 2 J_2  \nonumber \\
\beta_I &=&  -4 J_2 - 4 J_3  \; .
\end{eqnarray}

General cases of such Hamiltonians and their
equivalent surface formulations had been studied in some detail by Cappi {\it et.al.}
and found to have a very rich  phase structure
for generic choices of the couplings \cite{9a}, including
homogeneous, lamellar, disordered and bicontinuous phases. The Gonihedric model constitutes
a particular one-parameter slice in this space of Hamiltonians:
\begin{equation}
H = -2 \kappa \sum_{\langle ij\rangle }\sigma_{i} \sigma_{j}  +
\frac{\kappa}{2}\sum_{\langle \langle i,j\rangle \rangle }\sigma_{i} \sigma_{j} 
- \frac{1-\kappa}{2}\sum_{[i,j,k,l]}\sigma_{i} \sigma_{j}\sigma_{k} \sigma_{l}.
\label{e1}
\end{equation}
and for such a ratio of couplings $\beta_A=0$, which is the desired zero weight for plaquette area
suggested by translating the Hamitonian of equ.(\ref{steiner}). 
The
energy of the spin cluster boundaries for this Gonihedric model on a cubic
lattice is simply given by
\begin{equation}
E=n_2 + 4 \kappa n_4, 	
\end{equation}
where $n_2$ is the number
of links where two plaquettes on a spin cluster boundary meet at a right angle,
$n_4$ is the number of links where four plaquettes
meet at right angles, and $\kappa$ is a free
parameter which determines the relative
weight of a self-intersection of the surface.

The particular
ratio of couplings in equ.(\ref{e1}) also introduces a novel
semi-global symmetry into the Hamiltonian, as we discuss in the next section.
This is related
to a zero-temperature high degeneracy point where
it is possible to flip non-intersecting planes
of spins at zero energy cost.
This symmetry is further enhanced at $\kappa=0$ where the action becomes
a purely plaquette term
\begin{equation}
H =  {\bf -} \frac{1}{2} \sum_{[i,j,k,l]}\sigma_{i} \sigma_{j}\sigma_{k} \sigma_{l}.
\label{e2}
\end{equation}
and {\it any} plane of spins, even intersecting ones, may be flipped at zero energy cost
at $T=0$,
which suggests a highly degenerate ground state for the $\kappa=0$ Gonihedric model. Series expansions 
and cluster variational calculations, described below, show that the degeneracy 
of the Hamiltonian is  broken in the free energy for $\kappa \ne 0$, indicating a ferromagnetic low temperature phase in this case, but the nature of the order in the low temperature phase for $\kappa=0$ remains rather mysterious.

\section{Equilibrium Behaviour, by various means}

In what follows we use the standard definitions of quantities
such as the magnetisation
\begin{equation}
M = {1 \over L^3} \sum_i \sigma_i 
\label{ord}
\end{equation}
and the Binder cumulant
\begin{equation}
U = 1 - {\langle M^4 \rangle \over 3 \langle M^2 \rangle^2},
\end{equation}
in discussing critical properties,
where the $\langle \rangle$ brackets denote thermal averages.

The energy $\langle E \rangle$ is just given by the average of
the Hamiltonian,  
the specific heat $C$ by
the variance of the energy
and the susceptibility $\chi$ by the variance of the magnetisation.
The scaling exponents are then defined by the singular
properties of these quantities at the critical point(s) of the model
\begin{equation}
C = B + C_0 t^{-\alpha}, \; M = M_0 t^{\beta}, \; \chi = \chi_0 t^{-\gamma},
\; \xi = \xi_0 t^{-\nu}
\label{eF1}
\end{equation}
where $t = | (T - T_c) / T_c| $ is the reduced temperature, $\xi$ is the correlation length and
$\nu$ is the correlation length exponent, giving the divergence of the
correlation length at criticality. 

These can also be recast
as finite-size scaling relations
\begin{equation}
C = B' + C_0' L^{\alpha \over \nu}, \; M = M_0' L^{-\beta \over \nu}, \; \chi = \chi_0' L^{\gamma \over \nu}
\label{eF2}
\end{equation}
for lattices of linear extent $L$, which is  more convenient when $T_c$ is not known exactly, which is usually 
the case.

\begin{figure}
\centering
\includegraphics[height=5cm]{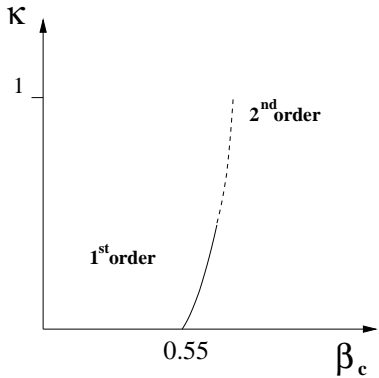}
%
%
\caption{The phase transition line as $\kappa$ is varied.}
\label{phase_kappa}       
\end{figure}

Various methods have been used to investigate the phase diagram 
of the Gonihedric model, ranging from generalised mean field 
techniques, through cluster variational calculations to Monte Carlo simulations.
The gross phase structure of the Gonihedric model was apparent even in the earliest
Monte Carlo work \cite{10}, and is shown schematically in Fig.\ref{phase_kappa}. For $\kappa=0$ and small 
values of $\kappa$ there is a first order transition \cite{11}, 
signified by a jump in the energy, $E$, at the transition point and
first order (volume) scaling in the specific heat.

As $\kappa$ was increased the transition softened to second order, but Monte Carlo 
simulations and other approaches have struggled to produce  
a consistent set of critical exponents. The first direct Monte-Carlo simulations \cite{10}, for instance, found $\nu = 1.2(1), \gamma=1.60(2), \beta=0.12(1)$ when $\kappa=1$, but 
later work using the scaling of the surface tension of
a spin interface \cite{12} obtained $\nu = 0.44(2)$ and $\gamma / \nu = 2.1(1)$.

Other approaches also gave results at variance with these:
a low temperature expansion by Pietig and Wegner found $\alpha = 0.62(3), \beta=0.040(2)$ and $\gamma=1.7(2)$
at $\kappa=1$ \cite{13}; whilst Cirillo {\it et.al.} \cite{14} found $\beta=0.062(3)$ and $\gamma=1.41(2)$ using a 
combination of the cluster variational method and Pad\'e approximants (``CVPAM'') \cite{15}.

These cluster variational calculations, ground state enumerations and mean field calculations  all 
use a similar starting point -- the decomposition of the lattice into elementary cubes
and we now discuss these methods in some detail since they are useful in sketching out the phase structure.
Indeed, the cluster variational method, allied with Pad\'e approximant techniques 
has succeeded in providing a very plausible picture of the Gonihedric phase diagram by
looking at an extended two-parameter Hamiltonian. 

\begin{figure}
\centering
\includegraphics[height=4.5cm]{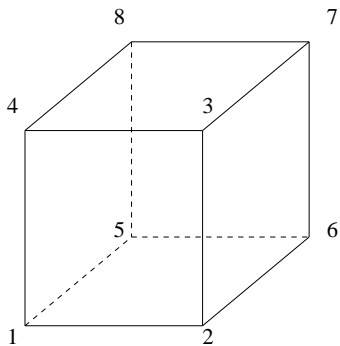}
%
%
\caption{The ground-state and mean field analyses of the Gonihedric model
require working with elementary cubes rather than single sites to capture
the various possible spin configurations.}
\label{cube}       
\end{figure}

One of the elementary cubes from the $3D$ cubic lattice is shown in Fig.\ref{cube}. 
The full lattice
Hamiltonian may be written as a sum over individual cube Hamiltonians
\begin{equation}
h_c = \frac{\kappa}{2}\sum_{\langle i,j \rangle} \sigma_{i} \sigma_{j} - \frac{\kappa}{4}  \sum_{ \langle \langle i,j
\rangle \rangle }\sigma_{i} \sigma_{j} 
+  \frac{1-\kappa}{4} \sum_{[i,j,k,l]}\sigma_{i} \sigma_{j} \sigma_{k} \sigma_{l} \; 
\end{equation}
which differs from the full Hamiltonian by the symmetry factors in front of each term. 
If the lattice can be tiled by
a cube configuration minimising the individual $h_c$
then the ground state energy density is
$E_0 = min\;  h_c$. 

We list some of the inequivalent spin
configurations on a single cube and their
multiplicities in  Table.1. 
\begin{table}
\centering
\caption{{\it Elementary cube ground state energies}}
\label{tab:1}       
\begin{tabular}{|c|c|c|c|c|}    \hline
State & Top  & Bottom& Energy & Multiplicity  \\ \hline
$\psi_0$ & {\tiny + +}& {\tiny + +} & $-3/2 -3 \kappa /2$ & 2 \\
       & {\tiny + +}& {\tiny + +} & &        \\  \hline
$\psi_{\bar 0}$ & - {\tiny +}& {\tiny +} - & $-3/2 +21\kappa /2$ & 2 \\
       & {\tiny +} - & - {\tiny +}& &   \\ \hline
$\psi_{6}$ &  - - &  {\tiny +  +} & $  -3/2 -3 \kappa /2$ & 6 \\
         &  - -   & {\tiny + + }  & &  \\  \hline
$\psi_{\bar 6}$ &  {\tiny + + } &  - -  & $  -3/2 +5 \kappa /2$ & 6 \\
         &  - -   & {\tiny + + }  & &  \\  \hline
\end{tabular}
\end{table}
In the list of spins the first column represents one face of the cube
and the second the other and the notation $\psi_0,...$ is borrowed from \cite{9a}
The antiferromagnetic image
of a configuration
is obtained by flipping the three nearest neighbours and the spin
at the other end of the cube diagonal from a given spin and is denoted by
an overbar.

With the Gonihedric values of the couplings there is a freedom to flip spin planes 
in the ground state
as $\psi_0$, which would represent a ferromagnetic state
when used to tile the lattice, and $\psi_6$ which would represent flipped
spin layers,
have the same energy for any value of $\kappa$. 
The degeneracy increase when
$\kappa=0$, the club
of
states of energy $-3/2$ is now composed of $\psi_0, \psi_{\bar 0}, \psi_6, \psi_{\bar 6}$.

This degeneracy means that the definition of an order parameter for the low-temperature phase is a moot point. 
The standard magnetisation $M$
will not serve at zero temperature as is clear from consideration of the ground state energies above, since the freedom to flip spin planes
means that it is identically zero. As we have noted, it appears that this symmetry is broken for $\kappa \ne 0$ at non-zero temperature,
but in Monte Carlo simulations the magnetisation is still indistinguishable from zero, presumably because of 
the difficulty of removing such flipped planes within the timespan of the simulation.
Even staggered
magnetizations do not suffice as order parameters as the interlayer
spacing between the flipped planes of spins can be arbitrary.

One possibility on a finite lattice is to
force the model into the ferromagnetic
phase with a suitable choice of
boundary conditions such as fixed boundary conditions. Although this
allows the use of a standard magnetic order parameter, it
does some violence to both the scaling properties and the ground state structure
structure itself. The corrections to first-order scaling, for example, are much stronger with fixed boundary conditions than with
the more familiar periodic boundary conditions,
\begin{table}[h]
\centering \caption[{\em Nothing.}] {{\em Scaling laws for
Periodic {\em versus} Fixed Boundary Conditions for the critical temperature, maxima of the specific heat and susceptibility and minimum of the Binder cumulant. In all cases a d-dimensional cubic lattice of length $L$
is considered.}} \vspace{3ex}
\begin{tabular}{|l|c|c|} \hline
    & P.B.C. & F.B.C. \\ \hline
  $\beta_{c}^{peaks}(L) =$ & $\beta_{c}(\infty)+ \frac{\theta_1}{L^{d}}+O(\frac{1}{L^{2d}})$ &
   $\beta_{c}(\infty)+ \frac{a_1}{L}+O(\frac{1}{L^{2}})$ \\
 $C_{max}(L) =$ & $\gamma_{0}+\gamma_{2}L^{d}+O(\frac{1}{L^{d}})$ &
   $c_{0}+c_{2}L^{d}+O(L^{d-1})$ \\
$\chi_{max}(L)= $ &
$\delta_{0}+\delta_{2}L^{d}+O(\frac{1}{L^{d}})$ &
   $e_{0}+e_{2}L^{d}+O(L^{d-1})$ \\
$U_{min}(L) = $ &
$\Phi_{0}+\frac{\Phi_{1}}{L^{d}}+O(\frac{1}{L^{2d}})$ &
$B_{0}+\frac{B_{1}}{L}+O(\frac{1}{L^{2}})$\\ \hline
\end{tabular}
\label{tab:scal-laws}
\end{table}
although simulations of the $\kappa=0$ model with fixed boundary conditions have given good agreement with
this scaling \cite{16}. 

The approach taken to enumerating the ground states by 
splitting the lattice into elementary cubes can also be applied
to mean field theory for the model and to extensions of mean field theory such as the cluster
variational method.
In the mean field approximation the spins
are  replaced by the average site magnetizations
and an entropy term introduced to give the free energy.
In the Gonihedric model 
the total mean field 
free energy may be written as the sum of elementary cube free energies $\phi(m_{C})$
\begin{eqnarray}
\phi{(m_{C})} &=& - \frac{\kappa}{2}\sum_{\langle i,j \rangle \subset C} m_{i} m_{j} + \frac{\kappa}{4}  \sum_{\langle \langle i,j \rangle \rangle \subset C }m_{i} m_{j}  
    -  \frac{1-\kappa}{4} \sum_{[i,j,k,l] \subset C} m_{i} m_{j} m_{k} m_{l} 
 \nonumber \\
&+& \frac{1}{16}
\sum_{i \subset C}[(1+m_{i})ln(1+m_{i})+ (1-m_{i})ln(1-m_{i})]
\end{eqnarray} 
where $m_{C}$ is the set of the eight magnetizations of the elementary cube.
This in turn gives a set of eight mean-field equations
\begin{equation}
\frac{\partial\phi(m_{C})}{ \partial m_{i}}_{(i=1 {\ldots} 8)} =0
\end{equation}
(one for each corner of the cube)
rather than the familiar single equation for the standard nearest neighbour Ising
action.

If we solve these equations iteratively we arrive at 
zeroes for a paramagnetic phase or various combinations
of $\pm 1$ for the magnetised phases on the 
eight cube vertices, and the mean field ground state
is then given by gluing together the elementary cubes consistently
to tile the complete lattice, in the manner
of the ground state discussion.  Carrying out this program gives a rather simple
mean field phase diagram for the Gonihedric model
with action equ.(\ref{e1}), with a single transition
from a paramagnetic phase
to a degenerate ``layered'' phase
that is pushed down to $\beta=0$ at large $\kappa$.
The low temperature phase is generically
of the $\psi_{0,6}$ type, apart from $\kappa=0$ where
we see a $\psi_{\bar 0, \bar 6}$ phase that is degenerate
with these. A  more sophisticated treatment using the cluster variational
method, described below, suggests that the layered phase has a 
slightly higher free energy than the ferromagnetic phase so the low temperature
phase is in fact the ferromagnetic one when the model is in the regime with
a continuous transition.

The cluster variation method, or CVM for short, is based
on a truncation of the cluster (cumulant) expansion
of the free energy density functional on which the variational
formulation of statistical mechanics is based \cite{16a}. 
Unlike mean field theory it generally locates rather accurately the
boundaries between different phases in complex phase diagrams and, 
using the recently proposed cluster variation--Pad\`e approximant
method  one can extract 
non-classical, precise estimates of the critical exponents.

For the Gonihedric model, it is again possible to use the cube approximation of the
CVM, in which the maximal clusters  are the elementary
cubic cells of our simple cubic lattice; for this approximation the
free energy density functional has the form 
\begin{eqnarray}
\phi [\rho_8] &=& {\rm Tr} (\rho_8 H_8) + \frac{1}{\beta} \Bigg[
{\rm Tr} {\cal L} (\rho_8) 
- \frac{1}{2} \sum_{\rm plaquettes} 
{\rm Tr} {\cal L} (\rho_{4,{\rm plaquette}}) \nonumber \\
&& + \frac{1}{4} \sum_{\rm edges}
{\rm Tr} {\cal L} (\rho_{2,{\rm edge}}) - \frac{1}{8} \sum_{\rm sites}
{\rm Tr} {\cal L} (\rho_{1,{\rm site}}) \Bigg],
\label{fcube}
\end{eqnarray}
where $H_8$ is the contribution of a single cube to the Hamiltonian, ${\cal L}(x) = x \ln x$, $\rho_\alpha$ with $\alpha
= 8$ (4, 2, 1) denotes the cube (respectively plaquette, edge, site)
density matrix, and the sums in the entropy part are over all
plaquettes (edges, sites) of a single cube. 

Since the cluster variation method can be viewed as a generalised mean field
theory, it is clear that it can give only classical
predictions for the critical exponents. In order to overcome this
difficulty, one can use the  cluster
variation--Pad\`e approximant method (CVPAM), which has proven to be a quite accurate
technique for extracting exponents. The
basic idea of the CVPAM is that, since the CVM gives for Ising-like
models very accurate results far
enough from the critical point, one can try to extrapolate these
results in order to study the critical behaviour. In order to
determine the critical exponent of an order parameter $M$, for example,
one calculates $M(\beta)$ with the CVM up to a temperature at
which the error can be estimated to be very small (typically of order
$10^{-5}$), and then constructs, by a simple interpolation, Pad\`e
approximants for the logarithmic derivative of $M(\beta)$: the pole
and the corresponding residue of each Pad\`e approximant are then
estimates for the critical temperature and for the critical exponent
respectively \cite{15}.

The CVPAM calculations give a phase diagram whose topology agrees
with Monte Carlo simulations \cite{14}. The value for the magnetic exponent is found to be
$\beta = 0.062 \pm 0.003$. 
Another important observation from the CVPAM calculations is
that at finite temperatures 
there is a violation of the flip symmetry 
of the Gonihedric Hamiltonian when $\kappa \ne 0$,  
so that in the ordered
region of the model  the ferromagnetic phase is always stable 
with respect to the lamellar phase. 

\begin{figure}
\centering
\includegraphics[height=5.5cm]{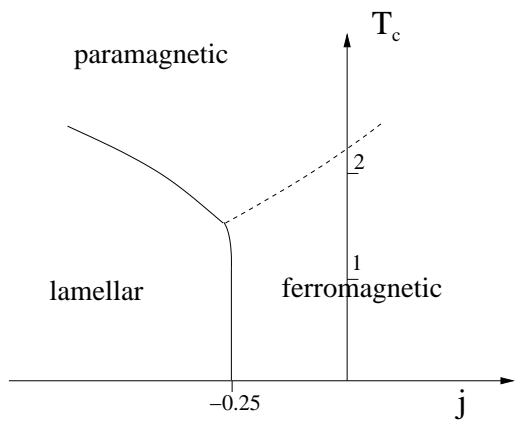}
%
%
\caption{A $\kappa=1$ slice of the two-parameter Hamiltonian phase diagram.}
\label{phase_j}       
\end{figure}

Applying CVPAM methods to a slight extension to the Gonihedric Hamiltonian
gives useful insight into the nature of the 
phase diagram \cite{17}. The Hamiltonian is modified to
\begin{equation}
H = - \sum_{\langle ij\rangle }\sigma_{i} \sigma_{j} 
 - j  \sum_{\langle \langle i,j\rangle \rangle }\sigma_{i} \sigma_{j} 
 - { 1 - \kappa \over 4 \kappa}  \sum_{[i,j,k,l]}\sigma_{i} \sigma_{j}\sigma_{k} \sigma_{l}.
\label{e4}
\end{equation}
with a second parameter $j$ which reduces to the Gonihedric model, with suitably rescaled
couplings, when $j = - 0.25$. The CVPAM calculations (and independent transfer matrix calculations \cite{18})
for this modified Hamiltonian give a phase diagram whose $\kappa=1$ slice is shown in Fig.\ref{phase_j}. 

The structure of this phase diagram gives some strong hints as to why the
critical exponent measurements in Monte Carlo simulations have been so inconsistent.
The dotted line in Fig.\ref{phase_j} is the usual $3D$ ferromagnetic Ising transition, along which 
the standard $3D$ Ising critical exponents such as $\nu= 0.6294$ would be observed. The line separating
the lamellar and ferromagnetic phases appears to be bent slightly to the left according to both 
CVPAM and transfer matrix calculations. The 
observed Gonihedric model transition 
will thus be strongly influenced by crossover effects close to the endpoint of
the standard Ising line which which is near, but not exactly at, $j=-0.25$. 

In such a case we would
expect the correlation length to diverge as
\begin{equation}
\xi \sim N^{\pm} t^{-\nu}
\end{equation}
where the new feature is the critical amplitude, which is given by
\begin{equation}
N^{\pm} \propto \Delta^{ ( - \dot \nu + \nu ) / \phi}
\end{equation}
with $\dot \nu$ and $\nu$ being the tricritical and $3D$ Ising values for the correlation length
exponent respectively, and  $\Delta = j + 0.25$. 
While CVPAM and transfer matrix calculations \cite{17,18} are both in agreement with this general picture 
the values obtained for the tricritical exponents and crossover exponent do not agree, the 
transfer matrix calculations finding $\dot \nu = 0.45(15)$ and $\phi=0.6(2)$, whereas
the CVPAM calculation finds $\phi=1.1(1)$. 
	
In summary, while the general features of the Gonihedric model equilibrium phase diagram are understood
both for $\kappa=0$ and $\kappa \ne 0$ and a plausible tricritical/crossover scaling scenario has been advanced to describe the observed values of the critical exponents,
these have still not been determined with any great degree of certainty. It is also difficult to define a suitable
magnetic order parameter for the low-temperature phase as the flip symmetry of the Hamiltonian means the
standard magnetisation will be zero in the ground state. Although both low temperature expansions
and CVPAM calculations of the free energy strongly suggest that this symmetry is broken for non-zero $\kappa$ at 
finite temperature, the standard magnetisation is still measured to be zero
in the low temperature phase in Monte-Carlo simulations for $\kappa \ne 0$. The flip symmetry appears
to be unbroken for the $\kappa=0$ (pure plaquette) case, so the low temperature phase for this
is highly degenerate.

A further difficulty 
which has been observed in attempting to extract critical exponents via Monte Carlo simulations is the very long 
autocorrelation times which the model exhibits, as was first remarked in \cite{18a}. These have the
effect of smearing out the observed critical singularities and rendering the extraction of
critical exponents problematic. 
With this in mind, we now turn to the non-equilibrium properties.

\section{Non-Equilibrium Behaviour, mostly by Monte Carlo simulations}

On heuristic grounds we would expect slow dynamics in the Gonihedric model since 
the destruction of a spin droplet will require an activation energy because 
of the weighting of any new edges in the Gonihedric Hamiltonian.
\begin{figure}
\centering
\includegraphics[height=7cm,angle=270]{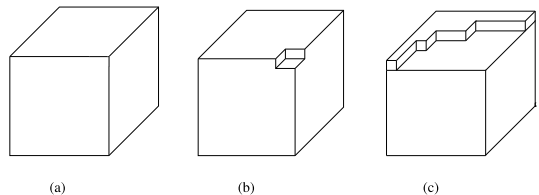}
%
%
\caption{Destroying a cubic excitation, by ``nibbling'' away the edges has an activation energy in the Gonihedric model.}
\label{nibble}       
\end{figure}
Indeed, when $\kappa=1$ the plaquette term drops out and one is left with
\begin{equation}
H = -2 \sum_{\langle ij\rangle }\sigma_{i} \sigma_{j}  +
\frac{1}{2}\sum_{\langle \langle i,j\rangle \rangle }\sigma_{i} \sigma_{j} \, .
\label{H_kappa1}
\end{equation}
which represents a particular case of a class of models 
with competing nearest neighbour and next-nearest neighbour interactions formulated
by Shore and Sethna specifically to investigate slow dynamics \cite{19}. We will denote such models
with generic coupling ratios Shore and Sethna models below. The Gonihedric Hamiltonian
with $\kappa=1$ displays some differences in its dynamics compared with the generic case.

Hand in hand with such slow dynamics and energy barriers we would also expect to see metastability in the model and
this is backed up by both CVPAM calculations and direct integration in Monte Carlo simulations
to obtain the free energy. For $\kappa=0$ the CVPAM calculations of the free energy show a region of metastability both above and below the first order transition point at $\beta = T^{-1} \sim 0.55$ \cite{14}.
\begin{figure}
\centering
\includegraphics[height=6cm]{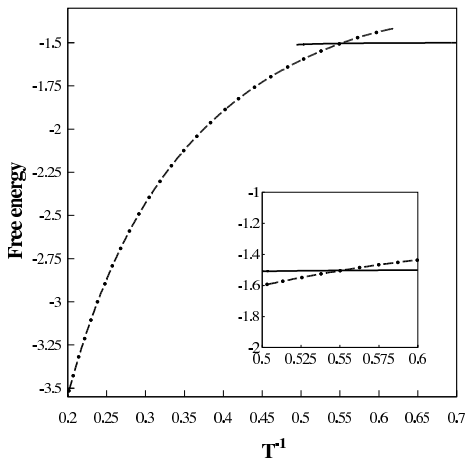}
%
%
\caption{The free energy from a CVPAM calculation when $\kappa=0$. The solid line is the low
temperature phase
and the dashed dotted line the high temperature paramagnetic phase.}
\label{free_kappa0}       
\end{figure}
To calculate the free energy of the model in a Monte Carlo simulation \cite{20} we  used the following
equations:
\begin{eqnarray}
\phi_{{\rm low T}} &=& E - T\int_{0}^{T} \frac{C}{T} \; dT, \nonumber \\ 
\phi_{{\rm high T}} &=&  -T s(\infty) + T\int_{0}^{1/T} E \; d(\frac{1}{T}).
\end{eqnarray}
In the above equation $C$ and $E$ denote the specific heat and internal energy, respectively 
(measured using the standard formulae in the Monte Carlo simulation), and $s(\infty)=\ln (2)$ is the entropy per site
at infinite temperature.
\begin{figure}
\centering
\includegraphics[height=6.5cm,angle=270]{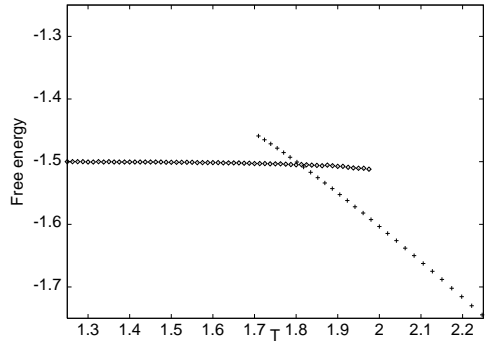}
%
%
\caption{The free energy from a Monte Carlo simulation when $\kappa=0$. The transition point and the region of metastability are both in excellent agreement with the CVPAM calculation of Fig.\ref{free_kappa0}.}
\label{free_int}       
\end{figure}

\begin{figure}
\centering
\includegraphics[height=6cm]{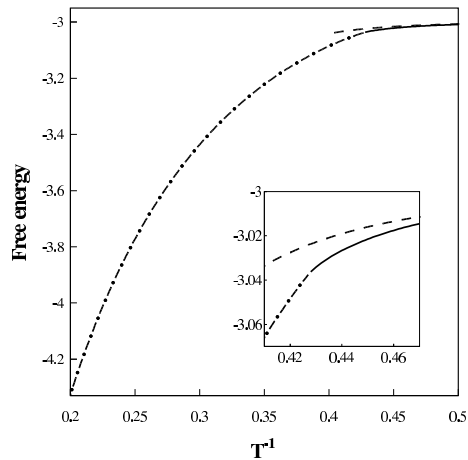}
%
%
\caption{The free energy from a CVPAM calculation when $\kappa=1$. The solid line is the ferromagnetic phase,
the dashed dotted line the high temperature paramagnetic phase and the dashed line the lamellar phase}
\label{free_kappa1}       
\end{figure}

We also find, see Fig.\ref{free_kappa1} and as advertised earlier, that the lamellar low temperature phase has a higher free
energy than the ferromagnetic phase when $\kappa=1$
indicating that the true low temperature phase is ferromagnetic in this case, although this does not
appear to be realised within the timescale of Monte Carlo simulations carried out so far.

Qualitative numerical experiments provide plenty of evidence for interesting
dynamical behaviour in the model, particularly with respect to quenches \cite{20}. In the $\kappa=0$ case
to study the evolution of a random configuration quenched to low temperature, one can measure the
energy excess $\delta E(t)=E(t)-E_0$ over the ground state energy $E_0$.
One expects that the inverse of this quantity sets the
characteristic length scale of the system $l(t)$, which roughly corresponds to the average size
of domains.
There is convincing evidence that the generic behaviour in many systems with nonconservative
dynamics and scalar order parameter (i.e., conditions which are satisfied here)
$l(t)$ increases asymptotically in time as $l(t)\sim t^n$ and $n=1/2$.
However, in some systems $l(t)$ is known to increase much more slowly in time,
even logarithmically $l(t)\sim {\rm log}(t)$.
These rather exceptional systems include some random (at the level of the
Hamiltonian) systems, and also the Shore  and Sethna models 
at sufficiently low temperatures.
It is the energy barriers developing in these systems during the coarsening which
cause such a slow increase of $l(t)$, so the Gonihedric model would also be a good candidate
for such behaviour.

\begin{figure}
\centering
\includegraphics[height=6cm,angle=270]{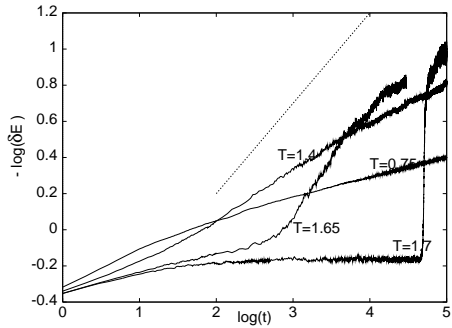}
%
%
\caption{The excess energy over the groundstate {\it vs} time during a quench to various temperatures.
The dotted line represents generic coarsening behaviour with an exponent of $1/2$.}
\label{deltaE}       
\end{figure}

The log-log plot of $1/\delta E(t)$ as a function of time for the Gonihedric model 
with $\kappa=0$ is shown in Fig.\ref{deltaE}
The presented results are obtained for $L=40$ but very similar behaviour was observed for
$L=30$.
From  Fig.\ref{deltaE} it is clear that for $T=0.75$ and $1.4$  the asymptotic slopes of the curves are much
smaller than 1/2 and there is a tendency for these curves to bend downwards.
Taking into account the absence of models with $n$ considerably smaller than 1/2 and the existence
of energy barriers in the Gonihedric model of basically the same nature as in the Shore and Sethna model, suggests that
for the examined temperatures the characteristic length asymptotically increases logarithmically in
time.
It would appear that  such a slow increase of $l(t)$ takes place even for $T=1.65$ and $1.7$,
but the behaviour of $l(t)$ for these temperatures is obscured by the metastability
effects, since before collapsing into the glassy phase the system 
remains in the liquid state for some time.

The difference between the Gonihedric model and the Shore and Sethna model becomes clear when 
we approach the lower boundary of the metastable region which we roughly estimate to lie at
$T=T_{{\rm l}} \sim 1.7$
by {\it increasing} the temperature. 
In the Shore and Sethna model for temperatures below the critical point but above a corner-rounding transition 
thermal fluctuations roughen corners of domains and energy barriers are no longer relevant.
Consequently, the standard coarsening dynamics with $n=1/2$ is restored and the system rapidly evolves
toward the low-temperature phase.
On the contrary, in the Gonihedric model for $1.7<T<1.95$ a random quench does not even
evolve toward the low-temperature phase but remains disordered.

\begin{figure}
\centering
\includegraphics[height=6cm,angle=270]{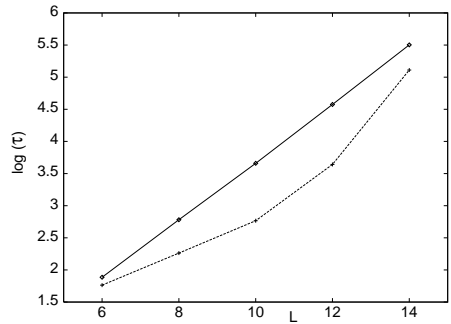}
%
%
\caption{The size dependence of the logarithm of the escape times $\tau_{{\rm liq}}$
($\Box$) and $\tau_{+-}$ ({\large +}).
Calculation of $\tau_{{\rm liq}}$ and $\tau_{+-}$ was done for $T=1.75$ and $T=1.8$, respectively.}
\label{f2}       
\end{figure}

The region of metastability can also be investigated through
measuring various characteristic times by imposing different initial and boundary conditions and
monitoring the evolution toward a final state.
To check the decay time associated with the stability of the high temperature ``liquid'' phase ($\tau_{{\rm liq}}$), 
a random initial configuration can be used
and then simulated at lower temperature until the energy reaches a sufficiently small value.
To estimate $\tau_{{\rm liq}}$ around 100 independent runs were necessary.
The results for $T=1.75$ are shown in Fig.\ref{f2} and they suggest that the escape time
increases at least exponentially with the system size, which is at first sight surprising for a 
model with finite range interactions. 

A later order of magnitude calculation \cite{21} for 
the size of a critical droplet (one that would grow and eventually take over the whole 
system with the stable phase) in the Gonihedric model by Swift {\it et.al.} showed that 
the results discussed here were for system sizes comparable or smaller than such a droplet, 
so the system size itself was setting the scale of energy barriers, leading to the observed exponential
growth in lifetimes. The energy barrier $\Delta$ to forming a droplet of size $R$ would be
\begin{equation}
\Delta = A \sigma R^{d-1} - B \delta f R^d	
\end{equation}
in $d$ dimensions, where $\sigma$ is the surface tension between stable and metastable phases,
$\delta f$ is the bulk free energy difference between the two phases and $A$ and $B$ are determined by
the geometry of the droplet. Maximising $\Delta$ gives the critical radius $R^*$
\begin{equation}
R^* = { A ( d -1 ) \sigma \over B \delta f d }.
\end{equation}
At low $T$ in $3D$ this gives a timescale for the nucleation of such 
a critical droplet of the form
\begin{equation}
\tau \sim \exp ( 4 A^3 \sigma^3 / 27 B^2 \delta f^2 T)	
\end{equation}
and inserting plausible estimates for $\sigma$ and $\delta f$ in the $\kappa=0$ Gonihedric model
gives  $R^* = 25$ for cubic droplets and $\tau \sim 10^{47}$ Monte Carlo steps.
F

To check the stability of the low temperature ``crystal'' phase, one should measure the time
($\tau_+$) needed for the crystal to be transformed into the liquid.
It would be particularly interesting to examine the size dependence of $\tau_+$ for $1.8<T<1.95$,
i.e., for temperatures where the crystal is metastable.
It was found, however, that this quantity increases very rapidly with the system size and in
this temperature range it is virtually impossible to increase the system size beyond $L=6$.
The stability of this phase might be also inferred from measurements 
of  the time ($\tau_{ +-}$) needed to shrink a cubic excitation of
size $L$.
This technique parallels that which has already been applied to the Shore and Sethna model: the
initial configuration has ``up '' spins at the boundary of the cube of size $L+2$ (which are kept
fixed) and ``down '' spins inside this cube.
Simulations are performed until the magnetisation of the interior of the cube decays to zero and the
time needed for such a run is recorded.
To calculate $\tau_{+-}$ at a given temperature 100 independent runs were again made.
The results for $T=1.8$ suggest that $\tau_{+-}$ increases approximately exponentially
with $L$, whereas
above the corner-rounding transition in the Shore and Sethna model one expects $\tau_{+-}\sim
L^2$, in which case the data in Fig.\ref{f2} would bend noticeably downwards.

\begin{figure}
\centering
\includegraphics[height=5cm]{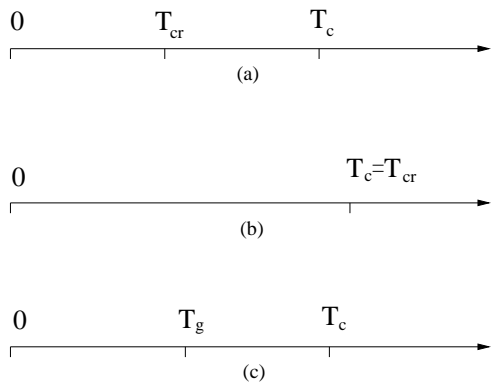}
%
%
\caption{The phaselines for the Shore-Sethna model {\bf (a)}, the Gonihedric model with a continuous transition {\bf (b)}
and the Gonihedric model with a first order transition (in particular $\kappa=0$) {\bf (c)}.}
\label{phaseline}       
\end{figure}

Other analogous numerical experiments may be carried out for non-zero $\kappa$ and the 
tentative picture which emerges \cite{22}
is shown in Fig.\ref{phaseline}.
The salient features are that generic Shore Sethna models with competing nearest neighbour and next nearest neighbour interactions have logarithmically slow low temperature dynamics, with a transition to
standard $t^{1/2}$ coarsening behaviour above  a corner rounding transition at $T_{cr}$. For the Gonihedric model
with a continuous transition ({\it e.g.} $\kappa=1$) simulations suggest that this corner rounding transition is
pushed close to (or may even be at) the equilibrium critical temperature $T_c$. The $\kappa=0$ Gonihedric model 
displays rather different properties. In this case there is a region of strong metastability on either side
of a first order equilibrium transition at $T_c$. We tentatively identify the lower boundary of the region 
of metastability with a glass transition temperature $T_g$. We now discuss why it is tempting to do so.

Glassy effects in simulations are usually discerned by looking at appropriate two-time quantities
such as the spin-spin autocorrelation
\begin{equation}
C ( t , t_w ) = \langle \frac{1}{N} \sum_i \sigma_i ( t_w) \sigma_i ( t + t_w)	\rangle
\end{equation}
where $N = L^3$ and $t_w$ is a {\it waiting time} before measurements are commenced.
Both normal coarsening and glassy dynamics
are expected to display strong waiting time dependence (``aging'') , but one may distinguish  the type I
aging dynamics seen in coarsening from the type II aging seen in glassy systems  by looking at
the overlap distribution function
\begin{equation}
Q ( t_w + t, t_w + t) 	= \langle \frac{1}{N} \sum_i \sigma^{(1)}_i ( t + t_w ) \sigma^{(2)}_i ( t + t_w ) \rangle \; .
\end{equation}
This is measured by relaxing the system from a disordered state for a time $t_w$
at which point
it is cloned into two sets of spins, $\{\sigma^{(1)}\}$ and $\{\sigma^{(2)}\}$, and evolved with independent random numbers for  a further time $t$. One finds that $Q ( t_w + t, t_w + t) \rightarrow 0$
as $C ( t , t_w ) \rightarrow 0$ in type II aging -- the heuristic idea being that no matter when the two
systems are separated their configurations  continue to move apart, which is suggestive of a truly rugged
free energy landscape as found in glasses. If this were not the case, as in type I aging, one would
find  $Q ( t_w + t, t_w + t) \rightarrow q$ for some constant $q$ as $ C ( t , t_w ) \rightarrow 0$.
For the $\kappa=0$ Gonihedric Ising model type II aging is clearly seen with these criteria
\cite{21,22,23}

A further glassy feature is the behaviour of energy autocorrelation \cite{21}
\begin{equation}
A ( t , t_w ) = \langle E ( t_w) E ( t + t_w ) \rangle	
\end{equation}
for $T_g < T < T_c$. It is found that   a stretched exponential form
\begin{equation}
A ( t , t_w ) \sim  A_0 \exp \left( - \left( { t \over \tau }\right)^{\beta} \right)
\end{equation}
fits this quite well,
where the relaxation time diverges as
\begin{equation}
\tau \sim { A \over T - T^* }	
\end{equation}
and the value for $T^*$ is very close to the measured value for $T_g$. The exponent $\beta$
is temperature dependent and has been measure to fall in the range $0.6-0.8$

\begin{figure}
\centering
\includegraphics[height=7cm,angle=270]{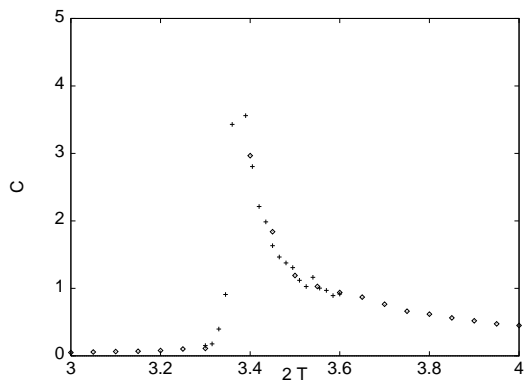}
%
%
\caption{The specific heat $C$ as a function of temperature $T$ calculated from the variance of
the internal energy for $L=24$ ($\diamond$) and $L=40$ ({\large +}).
At each temperature the system was relaxed for $10^3$ Monte Carlo steps and 
then measurements performed
during $10^4$ Monte Carlo steps.}
\label{heat}       
\end{figure}

Further numerical cooling experiments suggest that it is possible that the Gonihedric model with $\kappa=0$ may be displaying not just glassy behaviour,
but {\it ideal} glassy behaviour \cite{24}. Some time ago Anderson proposed that the glassy transition, which is a kinetic
phenomenon, might be related in the limit of vanishingly small cooling rate with a thermodynamic
transition \cite{25}.
It is found that
for the Gonihedric Ising model the peak in the specific heat of the liquid
occurs  at, or very close to, the temperature at which the internal energy jumps under very slow cooling.
The specific heat shown in Fig.\ref{heat}  is a thermodynamic quantity.
It was calculated in a (standard) quasi-equilibrium manner: after fixing a temperature we relaxed
the system and then measured the variance of the internal energy.
The sharp peak seen in  Fig.\ref{heat} indicates a thermodynamic-like singularity in this model.

\begin{figure}
\centering
\includegraphics[height=7cm,angle=270]{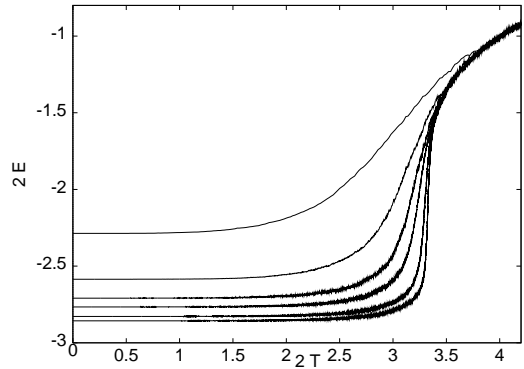}
%
%
\caption{The internal energy $E$ as a function of temperature for (from the top) $r=$ 0.02, 
0.002, 0.0005, 00002, 0.00005 and 0.00002.}
\label{Ujump}       
\end{figure}

On the other hand an almost vertical drop of the internal energy under continuous cooling, shown
in Fig.\ref{Ujump}, indicates the dynamic, glassy transition.
The results were obtained by simulating the  model under continuous cooling with a constant cooling rate $r$ and initial
temperature $T_0=2.1$ ($T_0>T_{{\rm g}}$).
This means that temperature as a function of time is given by $T=T_0-rt$.
Calculations were performed for several system sizes $L$ in order to ensure that $L$ was sufficiently
large.
For example for $r=0.02$ the system size $L=30$ is sufficient to obtain size-independent results but for
$r=0.00002$ we had to take $L=70$.
One can see that although $r$ decreases by three decades, the zero-temperature energy  
approaches the ground-state energy very slowly
and that lowering the cooling rate $r$, sharpens the transition.
At first sight one might expect that in the limit $r\rightarrow 0$ the transition becomes
infinitely sharp and coincides exactly with thermodynamic singularity at the peak of the specific 
heat. However, as we have seen, the metastability of
the liquid is a finite (but large) time/size effect and neither the peak nor the internal energy drop can
be made perfectly sharp.

Anderson's idea has had a rather limited experimental support.
The main problem is that under slow cooling real liquids do not get trapped in the
glassy phase but instead tend to crystallise.
The reason for that is that when liquid is cooled below the melting point it becomes metastable and
within a finite time due to heterogeneous or homogeneous crystal nucleation it
then crystallises.
Only under sufficiently fast cooling can the crystal nucleation be avoided and the liquid
trapped in the glassy state.
In this context, the Gonihedric model with plaquette interactions appears to correspond to an almost ideal 
glass with an extremely large lifetime for the metastable state.

\begin{figure}
\centering
\includegraphics[height=3.5cm]{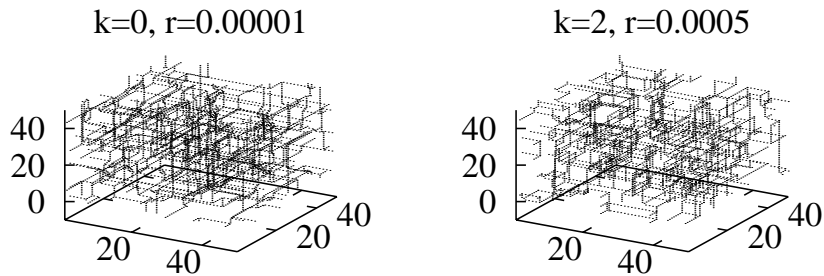}
%
%
\caption{A distribution of unsatisfied plaquettes in the zero-temperature glassy phase.
Simulations were made for the system size $L=100$.}
\label{frust}       
\end{figure}

A final subject which merits further investigation is the nature of the self-induced disorder in the glassy phase.
To get some insight into the $\kappa=0$ case it is possible to
look  the distribution of
unsatisfied plaquettes in the glassy phase (i.e., plaquettes
contributing energy above the ground state) \cite{26}.
A random high-temperature sample was slowly cooled down to zero temperature
and  the final configuration was used to locate unsatisfied plaquettes and
their spatial distribution is shown in Fig.\ref{frust}.
For comparison we also present similar simulations for the $\kappa=2$ case.
One can see that in both cases the energy from these unsatisfied plaquettes is concentrated in linear segments
(around $90 \%$ in both cases for slow cooling).
Interestingly, low temperature excitations observed in the $3D$ Edwards Anderson {\it spin} glass
\begin{equation}
H = - \sum_{\langle ij \rangle} J_{ij} \sigma_i \sigma_j	
\end{equation}
where the nearest neighbour couplings $\{ J_{ij} \}$ are quenched Gaussian random variables
with zero mean and unit variance, appear to have a rather similar stringy structure \cite{27}.
It is possible that the 
self-induced disorder in the $\kappa=0$ Gonihedric model is giving rise to similar characteristics 
for the excitations as the
quenched disorder which is imposed {\it a priori} in the finite-dimensional spin glass.

A further point of contact with spin glasses is that a spin glass version of the 
$\kappa=0$ Hamiltonian (still on a $3D$ cubic lattice)
\begin{equation}
H = - J_{\square} \sum_{[i,j,k,l]} \sigma_i \sigma_j \sigma_k \sigma_l	\; ,
\end{equation}
where the $J_{\square}$ are now quenched random variables, was also found 
to display numerous common properties with structural glasses \cite{27a}. Many
of the observed numerical features seen in simulations of this model, such as the stretched
exponential correlations and divergence of the autocorrelation time are extremely similar to those
discussed here for the $\kappa=0$ Gonihedric model. The inference from this might be 
that the additional quenched disorder arising from the $J_{\square}$ may be gratuitous, since
the plaquette Hamiltonian on its own is enough to give the same, or very similar, behaviour.

Self-induced disorder has been posited before as a mechanism for glassy behaviour in 
various deterministic mean-field models. One well-known example is the Hamiltonian originally formulated by Golay
\cite{27b}
\begin{equation}
H = - { J_0^2 \over 2 N} \sum_{k=1}^{N-1} \left[ \sum_{i=1}^{N-k} \sigma_i \sigma_{i+k} \right]^2
\label{Golay}
\end{equation}
in the context of obtaining binary sequences with small autocorrelations. It was shown that 
many of the properties of the deterministic Hamiltonian in equ.(\ref{Golay}) were reproduced by
the disordered Hamiltonian
\begin{equation}
H =  - { 1 \over 2 N } \sum_{k=1}^{N-1} \left[ \sum_{i,j=1}^N J^{(k)}_{ij} \sigma_i \sigma_j \right]^2
\end{equation}
where the $J^{(k)}$ were random connectivity matrices \cite{27c}. The contrasting feature 
for the Gonihedric model is that the self-induced disorder is appearing with finite range couplings.

The non-equilibrium, dynamical properties of the Gonihedric Ising model thus display
many interesting features
both for $\kappa=0$ and $\kappa \ne 0$. For the case of $\kappa=0$ (the purely plaquette action)
there are clear parallels with the observed behaviour of real glass formers, so it offers an appealing
lattice model for investigations of glassy dynamics by virtue of its simplicity. Indeed, there may also be similarities
with finite dimensional spin glasses, as we have seen above.

\section{Variations on the Glassy and Gonihedric  Themes}

There are a plethora of other lattice models which can be formulated 
to display glassy properties \cite{28,29}. In many cases, these have trivial static Hamiltonians
of the form
\begin{equation}
H = \sum_{i=1}^N n_i,
\label{Hamiltonian}
\end{equation}
where the $n_i = 0,1$ are defined on each site of
a cubic lattice of linear size $L$, which has periodic boundary
conditions and can be though of as labelling the (im)mobility of a 
site. The interesting (glassy) behaviour is induced in these cases via constrained dynamics.

Other models may possess a non-trivial Hamiltonian, but a trivial equilibrium phase diagram.
Nonetheless, the dynamical properties of such models may still be of interest.
The 2D Gonihedric Ising model is one such case
\begin{equation}
\ham_{gonih}^{\textit{\tiny 2D}}=
-\kappa\sum_{<i,j>}\sigma_i\sigma_j 
+\frac{\kappa}{2}\sum_{\ll i,j\gg}\sigma_i\sigma_j
-\frac{1-\kappa}{2}\sum_{[i,j,k,l]}\sigma_i\sigma_j\sigma_k\sigma_l \; . 
\nonumber 
\end{equation}
where the form is similar to the $3D$ model, but the relative weights of the terms differs.
This is constructed to weight the corners of $2D$ spin clusters, rather than their edge length,
which is the effect of the standard $2D$ Ising action.
Via a mapping to the 8-vertex model it is possible to show that the $2D$ Gonihedric Hamiltonian does
not display a static phase transition. Just as for the 
trivial Hamiltonian models with  constraints, the dynamics displays many interesting
features \cite{30}.

We have concentrated exclusively on Ising spins here. It is also possible to define a 3D Gonihedric Potts models
\begin{eqnarray}
H &=&- 2 \kappa \sum_{x,\mu>0} (2
\delta_{\sigma_x,\sigma_{x+\mu}}-1) + \frac{\kappa}{2}\sum_{x,\mu \ne
\nu, \mu>0} (2 \delta_{\sigma_x,\sigma_{x+\mu+\nu}}-1) \nonumber
\\ &-& \frac{1- \kappa}{2}\sum_{x,\mu \ne \nu, \mu>0, \nu>0} (2
\delta_{\sigma_x,\sigma_{x+\mu}}\delta_{\sigma_{x+\nu},
\sigma_{x+\mu+\nu}}-1) \cdot \nonumber \\ && \hspace*{3.2cm} (2
\delta_{\sigma_x,\sigma_{x+\nu}}\delta_{\sigma_{x+\mu},
\sigma_{x+\mu+\nu}}-1) \cdot \nonumber \\ && \hspace*{3.2cm}(2
\delta_{\sigma_x,\sigma_{x+\mu+\nu}} \delta_{\sigma_{x+\nu},
\sigma_{x+\mu}}-1) \label{hamil1}
\end{eqnarray}
where the spin variables $\{ \sigma \}$ now take on $q$
values \cite{31}.
In this case the model appears to retain a first order equilibrium transition for larger $\kappa$ than with Ising spins, but the dynamical properties of the model remain to be investigated.

The idea that multi-spin, specifically plaquette, interactions  give rise to self-induced
disorder might also be applied to models with continuous spins such as gauge glasses. The standard $3D$ 
gauge glass Hamiltonian is of the form \cite{32}
\begin{eqnarray}
H = - \sum_{<i,j>} \cos \, ( \theta_i - \theta_j + A_{ij} )
\end{eqnarray}
where the  $A_{ij}$ are quenched, random variables uniformly distributed in $[ 0, 2 \pi ]$. 
A plaquette interaction of the form
\begin{eqnarray}
H = - \sum_{[i,j,k,l]} \cos \, ( \theta_i - \theta_j + \theta_k - \theta_l )
\end{eqnarray}
would presumably give rise to a similar sort of frustration
as the quenched disordered $\{ A_{ij} \}$.

A final avenue for further research might be the dual model to the $\kappa=0$ 
Gonihedric action, which may be written in several equivalent forms
including
\begin{equation}
H_{dual} = \sum_{\xi} \Lambda^{\chi} (\xi) \Lambda^{\chi} (\xi + \chi)
+ \Lambda^{\eta} (\xi) \Lambda^{\eta} (\xi + \eta)
+ \Lambda^{\zeta} (\xi) \Lambda^{\zeta} (\xi + \zeta)
\label{dual1}
\end{equation}
where $\Lambda^{\chi} = ( 1, 1, -1, -1)$, $\Lambda^{\eta} = ( 1, -1, 1, -1)$
and $\Lambda^{\zeta} = ( 1, -1, -1, 1)$ are one dimensional
irreducible representations of the fourth order Abelian group
and $\xi, \eta, \zeta$ are unit vectors in the dual lattice.
The spins may also be considered as Ising ($\pm 1$) spins
if we set $\Lambda^{\zeta} = \Lambda^{\chi} \; \Lambda^{\eta}$,
which gives the following Hamiltonian
\begin{equation} 
H_{dual} = \sum_{\xi} \Lambda^{\chi} (\xi) \Lambda^{\chi} (\xi + \chi)
+ \Lambda^{\eta} (\xi) \Lambda^{\eta} (\xi + \eta) 
+ \Lambda^{\chi} (\xi) \Lambda^{\eta} (\xi) \Lambda^{\chi} (\xi + \zeta)
\Lambda^{\eta} (\xi + \zeta).
\label{dual2}
\end{equation}
This is recognizable as a strongly anisotropically coupled Ashkin-Teller model. It is possible
that consideration of this dual model might shed some light on the definition of an
order parameter in the original Hamiltonian and also suggests an intriguing link with
{\it anisotropic} scaling properties.

\section{Endpiece and Acknowledgements}

Gonihedric $3D$ Ising models have had an interesting past and display  
novel properties both in equilbrium and dynamically.
The $\kappa=0$ (plaquette) Gonihedric Hamiltonian
appears to provide a simple lattice model with similar behaviour to structural 
glass formers, which also has some connections with spin glasses in spite of the absence of quenched disorder. 
Even after some fairly extensive investigation by Monte Carlo simulations and other means
over recent years, many features of the models still remain in need of clarification. They may well have an interesting future ahead of them too.

DAJ would like to thank G. Savvidy for an elementary Greek lesson, and for many discussions
over the years on the Gonihedric model. DAJ was partially supported by EU RTN-Network `ENRAGE': {\em Random Geometry
and Random Matrices: From Quantum Gravity to Econophysics\/} under grant
No.~MRTN-CT-2004-005616. Many of the results described here were obtained in collaboration with D. Espriu, 
E. Cirillo, G. Gonnella and A. Pelizzola who have all contributed to the understanding of the model.

\end{document}